\documentclass[useAMS,usenatbib]{mn2e}
\usepackage{natbib}
\usepackage{graphicx, setspace, subfig, latexsym, amssymb, amsmath, booktabs, amsfonts, wasysym, paralist}
\usepackage{multirow}
\usepackage{gensymb}
\usepackage{mathrsfs}

\title[Transient search of high-latitude Parkes pulsar survey]
{A search for rotating radio transients and fast radio bursts in the Parkes high-latitude pulsar survey}
\author[A. Rane et al.]{A. Rane$^{1}$, D. R. Lorimer$^{1,2}$, S. D. Bates$^{2}$, N. McMann$^{1}$, 
M. A. McLaughlin$^{1}$ \newauthor and K. Rajwade$^{1}$\\
$^{1}$Department of Physics \& Astronomy, West Virginia University, Morgantown, WV, 26506 USA\\
$^{2}$National Radio Astronomy Observatory, PO Box 2, Green Bank, WV 24944, USA
}
\begin{document}

\date{\today}

\pagerange{\pageref{firstpage}--\pageref{lastpage}} \pubyear{2015}

\maketitle

\label{firstpage}
\begin{abstract}
Discoveries of rotating radio transients and fast radio bursts (FRBs)
in pulsar surveys suggest that more of such transient sources await
discovery in archival data sets. Here we report on a single-pulse search for
dispersed radio bursts over a wide range of Galactic latitudes ($|b| <
60\degree$) in data previously searched for periodic sources by Burgay
et al. We re-detected 20 of the 42 pulsars reported by Burgay et al. and 
one rotating radio transient reported by Burke-Spolaor.
No FRBs were discovered in this survey. Taking into account this result, and other recent 
surveys at Parkes, we corrected for
detection sensitivities based on the search software used in the
analyses and the different backends used in these surveys and find that the 
all-sky FRB event rate for sources with
a fluence above 4.0~Jy ms at 1.4~GHz to be ${\cal R} =
4.4^{+5.2}_{-3.1} \times 10^3$~FRBs~day$^{-1}$~sky$^{-1}$, where the
uncertainties represent a $99\%$ confidence interval. While this rate
is lower than inferred from previous studies, as we demonstrate, 
this combined event rate is consistent with the results of
all systematic FRB searches at Parkes to date and does
not require the need to postulate a dearth of FRBs at intermediate
latitudes.
\end{abstract}


 \section{Introduction}\label{sec:intro}

The radio transient sky contains a number of known and potential
classes of sources which emit on timescales ranging from 
nanoseconds to years. These classes include the
Sun, planets, brown dwarfs, flare stars, X-ray binaries, ultra-high
energy particles, magnetars, $\gamma$-ray bursts,
maser flares, active galactic nuclei, radio supernovae, pulsars, annihilating
black holes, and transmissions from extraterrestrial civilizations (for a review, see,
e.g., \citealt{cordes2004}, \citealt{lazio2012}). Recently \citet{pietka2015} have analyzed the variability timescales of radio flares from nearly 90 different objects/events varying from flare stars to supermassive black holes in active galactic nuclei and have demonstrated that variability timescales could be used as an early diagnostic of source class in future radio transient surveys.

Within the last decade, as part of the analyses of pulsar surveys, a new
transient phenomenon --- known as fast radio bursts (FRBs) --- has
been identified. FRBs are categorized as short duration (few ms)
bursts that are non-repeating with dispersion measures higher than the
Galactic dispersion measure (DM) expected along that direction
\citep{cordesne2002}. The DMs range from 300--1600~cm$^{-3}$~pc
and have so far only been observed in the
1--2~GHz band. To date, 11 FRBs have been published\footnote{For an updated
  list, see http://astro.phys.wvu.edu/FRBs}
(\citealt{lorimerburst2007}; \citealt{keanefrb2012};
\citealt{thorntonfrb2013}; \citealt{thornton2013};
\citealt{spitlerfrb2014}; \citealt{burkespolaorfrb2014};
\citealt{petrofffrb}; \citealt{ravifrb2014}). Although, most of
these were discovered in archival surveys, FRB~140514
\citep{petrofffrb} and FRB~131104 \citep{ravifrb2014} are
real-time discoveries with a transient pipeline developed at the
Parkes telescope. Despite a significant amount of follow-up observations
which closely followed FRB~140514 \citep{petrofffrb}, and inspection
of archival surveys for other FRBs, none of
the FRBs currently known are associated with any source counterpart at other wavelengths.

As remarked by many authors, the high DMs of FRBs are suggestive of 
an extragalactic origin. Among the proposed extragalactic
source populations for FRBs are annihilating black holes
\citep{keanefrb2012}, flaring magnetars \citep{popovpostnov2013}, binary white dwarf mergers \citep{kashiyama2013}, binary neutron star mergers \citep{totani2013}, collapsing neutron stars
\citep{falcke&rezzolla2014}, and neutron star-black hole mergers \citep{lipunov2014}.

In view of the lack of any extragalactic counterparts identified
so far, a number of other scenarios remain equally intriguing.
\citet{loeb2014} considered the case of nearby Galactic flaring
main-sequence stars within 1~kpc as the sources of
FRBs. They propose that the excess dispersion comes from propagation
through the stellar corona. However \citet{luan2014} argued that the
free-free absorption would conceal any radio signal emitted from below
the corona. Also, FRBs exhibit quadratic dispersion curves that are consistent with the assumption of weak dispersion in a low density plasma. So, for the above model, in which the dispersion is concentrated in a relatively high density region, the quadratic dispersion approximation breaks down as the plasma frequency is comparable to the propagation frequency, posing significant problems for this model \citep{dennison2014}. However, \citet{maoz2015} have recently found possible flare stars in additional FRB fields using time-domain optical photometry and spectroscopy. The authors have evaluated
the chance probabilities of these possible associations to be in the range $0.1\%$ to
$9\%$. FRB~140514 was discovered in the radio follow-up observations of FRB~110220, three years apart within the same radio beam. \citet{maoz2015} also claim that these two FRBs are from the same repeating source with $99\%$ confidence and are consistent with the flare-star scenario with a varying plasma blanket between bursts. More FRB detections in general are necessary to confirm or refute a Galactic origin of FRBs.

Although FRBs have so far been observed over a range of Galactic
latitudes, their true distribution on the sky remains unclear.
Recently, based on an analysis of Parkes High Time Resolution Universe
(HTRU) mid-latitude survey data, \citet{petroffhtru2014} proposed that
there is a deficit of FRBs at intermediate latitudes. Further analyses
of archival and current surveys are required to investigate this
issue. In addition, \citet{keanepetroff2015} have assessed the
commonly used search algorithms used for FRB searches which impact the
 FRB sensitivities in individual surveys. Motivated by these
results, in this paper we present a single-pulse search which is
sensitive to both RRATs \citep{maurarrats2006} and FRBs on archival data previously searched
for pulsars at high Galactic latitudes by \citet{burgay2006}. In \S~2
we outline the basic methodology behind the search and the detection
criteria. Our results and single-pulse statistics of known pulsars re-detected in this search are discussed in \S~3 . In \S~4 we use our non-detection of FRBs to constrain their all-sky event rate. Finally, in \S~5 we summarize our results and present our conclusions.
\section{Search methods and Analysis}\label{sec:singlepulse}
\subsection{Survey parameters}

The Parkes high-latitude (PH) survey \citep{burgay2006} was designed
to find millisecond pulsars and exotic binaries which migrate away
from the Galactic plane. A total of 42 pulsars were detected in this
survey, of which seven were millisecond
pulsars and 18 were new discoveries. The analysis of the data by \citet{burgay2006} was carried
out using the standard periodicity search methods to find
periodic signals. However no single-pulse searches have been published
on these data. The single-pulse search method is very effective in
detecting sporadic sources like some pulsars (e.g.~nulling pulsars) and
RRATs in the time domain since these
might not be detectable in the standard periodicity searches
\citep{cordesmc2003}. Moreover, FRBs can of course only be detected through single-pulse searches (see below \S~2.2.2).

The PH survey covered a strip of the sky enclosed by Galactic
longitudes $220^{\degree} \leq l \leq 260^{\degree}$ and Galactic
latitudes $|b| \leq 60^{\degree}$ corresponding to a total sky
coverage of 3588~deg$^2$ in ~475 hours.
The survey began in November 2000 and ended in
December 2003 and made use of the 13-beam receiver on the Parkes 64-m
radio telescope. Data were collected simultaneously by 13 beams at a central
frequency of 1374~MHz with 96 frequency channels,
each 3~MHz wide. Each of the 6456 pointings was observed for 265~s. 
For more details about the receiver system and data acquisition, see \citet{burgay2006}.

\subsection{Single-pulse search method}\label{sec:sspmethod}

Each of these beams from the survey was processed independently
using the \textsc{sigproc} software 
package\footnote{http://sigproc.sourceforge.net}. The steps included
in our analysis are:
\begin{enumerate}
    \item dedisperse the raw data file at a range of trial DM values 
    and remove radio frequency interference (RFI) at zero DM; 
    \item search for individual pulses in the time series above 
    signal-to-noise (S/N) of five and with different widths;
    \item apply the detection criteria to filter in terms of DM, S/N,  
    and number of beams;
    \item manually inspect the resulting diagnostic plots.
\end{enumerate}
We describe each step in detail below, and give the appropriate
\textsc{sigproc} modules used.

\subsubsection{Dedispersion}

Radio signals are affected by interstellar dispersion, and as a result,
the higher frequencies of the signal traveling faster through the
interstellar medium arrive earlier than their lower frequency
counterparts. The time delay between the two frequencies $f_{1}$ and
$f_{2}$, (see, e.g., \citealt{psrhandbook}) is
\begin{equation}
\Delta t \simeq 4150 \mathrm{s} \left[
\left(\frac{f_{1}}{\rm MHz}\right)^{-2} -
\left(\frac{f_{2}}{\rm MHz}\right)^{-2}
\right] 
\left(\frac{\rm DM}{\rm cm^{-3}~pc}\right),
\label{timedelay}
\end{equation}
where DM is the integrated number density of free electrons along the
line of sight. This dispersion allows us to distinguish between
astrophysical and terrestrial signals. It causes a quadratic sweep across the band
 and may be removed by appropriately shifting the
frequency channels. Each time series was dedispersed over a range of
trial DM values using the {\tt dedisperse} routine in
\textsc{SIGPROC}. For this analysis, we have searched DMs in a range
from $0-10^4$~cm$^{-3}$~pc. The wide range of DM makes the search
sensitive to events that are highly dispersed. The trial DM step sizes used in the analysis were
calculated using an algorithm, {\tt
  dedisperse\_all}, originally described by
\citet{levin2012} which accounts for the amount of pulse broadening
caused by the size of the previous DM step and then determines the next trial DM.
A total pulse width
smearing due to the DM step in comparison to the value at the last DM is chosen to be $25\%$ (see, \S~Appendix). 
The total number of DMs searched was 249, as chosen
optimally by this program. For the DM steps used in our analysis, the average S/N loss is $\sim 1.5\%$ for DMs $<2000$~cm$^{-3}$~pc and for DMs between $2000-10000$~cm$^{-3}$~pc, the average S/N loss is $\sim 2.5\%$, calculated using Equations 12 and 13 of \citet{cordesmc2003}. 
The {\tt dedisperse} routine uses Equation 1 to
calculate the time delays for each test DM and applies to
frequency channels and the samples from each channel are then
averaged to form a dedispersed time series. 
In addition to this, it also performs zero-DM subtraction 
\citep{eatough2009} on the time 
series to remove any RFI at zero DM.

\subsubsection{Single-pulse search}\label{singlepulse}
Each dedispersed time series corresponding to a particular trial DM
was searched for transient events of different widths via the matched
filtering technique for top-hat pulses implemented in the program {\tt
seek}. This simple algorithm, which is an
implementation of the method described in \citet{cordesmc2003},
saves individual events that deviate by five
standard deviations from the mean of the time series. A
number of adjacent samples are added to search for pulses of different
widths. Each time series was
smoothed 15 times, corresponding to a maximum smoothing of $2^{15}$
times the sampling interval, i.e. pulse widths out to
4.096~s. If a pulse is detected in more than one of the
smoothed times series, only the highest S/N value is recorded. Following \citet{burkeIII2011}, the 
number of independent trials 
\begin{equation}
N = N_{\rm DM} \, \sum^{\rm j_{\rm max}}_{j=0} \frac{N_{\rm samp}}{2^j},
\end{equation}
where $N_{\rm DM}$ is the number of DM trials, $N_{\rm samp}$ is the number of samples in each dedispersed time series, and $j_{\rm max}$ corresponds to number of matched-filter widths used. We find $N = 1.1 \times 10^9$ for our observation. For
more details of this search method, see \citet{cordesmc2003}. 

\cite{keanepetroff2015} have assessed the performance of the search
algorithms commonly being used to discover FRBs. The authors point out some 
important concerns where sensitivity to FRBs is often unnecessarily reduced and that 
the single-pulse search routines within the {\tt dedisperse\_all} and {\tt
  seek} packages were less efficient compared to {\tt Heimdall} and {\tt destroy} packages \citep[see, e.g.,][]{petroffhtru2014}. Based on the simulations which used an injected signal of known strength,
\cite{keanepetroff2015} demonstrated that the recovered S/N for {\tt
  dedisperse\_all} and {\tt seek} was a function of pulse phase. We 
apply these results into our analysis, as described in \S~4.

\subsubsection{Detection criteria}

\begin{figure*}
\centering
\includegraphics[scale=0.37]{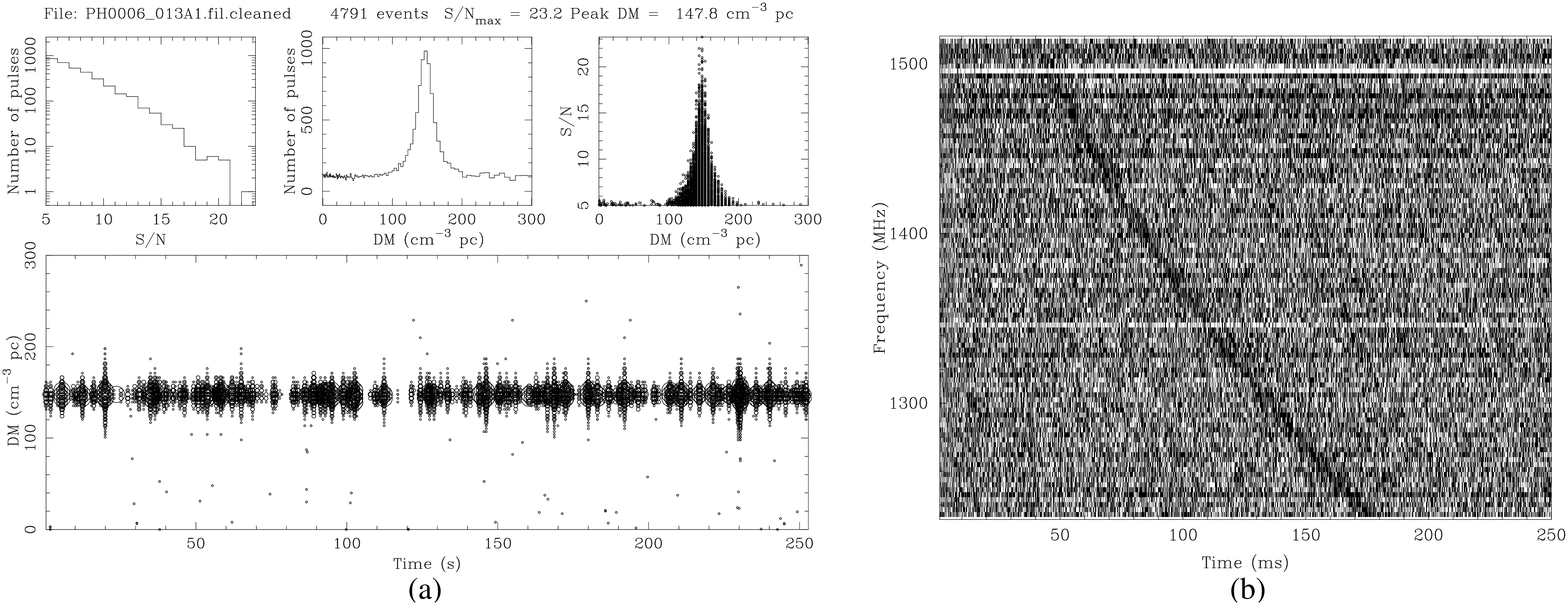}
     \caption{Example output of one single-pulse
       search processing pipeline, showing bright single pulses from
       PSR~J0837$-$4135 around a DM value of 147 ~cm$^{-3}$~pc. The top left panel
       shows the S/N distribution of the detected pulses, number of pulses
       versus trial DM (in top center), and S/N as a function of trial
       DM. The lower plot shows S/N of events as a function of time
       and trial DM. The size of the circles is linearly proportional
       to the S/N of each pulse. The dispersive
       delay in the frequency versus time plot is seen on the right for 250 milliseconds of data.}
     \label{fig:j0837-4135example}
\end{figure*}

These dispersed pulses can 
be displayed graphically using the {\tt plotpulses} program and the
resulting single-pulse plot can be seen in the left panel of
Fig. ~\ref{fig:j0837-4135example}. In the DM histogram, we
are able to detect sources that emit many weak pulses. Such sources might
not be detectable in the bottom plot. Conversely, the sources that emit only a
few strong pulses may only be detectable in the bottom plot. Often a
peak in the DM histogram at low DMs is seen which is indicative of
RFI. In some cases, RFI is seen at all DMs for a certain time
range which can further limit our ability to detect a transient
event. This would happen if there is a strong source of RFI that causes the receiver to saturate or some other local interference that is so strong that it shows up at all DMs for that time range. The single-pulse plots thus obtained were inspected manually (e.g. by searching for a
well-defined peak in the S/N versus DM plot, see 
Fig.~\ref{fig:j0837-4135example}). For manual inspection, we restricted the S/N threshold to six in order to keep the number of potential candidates at a manageable level.

The event detected by \citet{lorimerburst2007} was detected
simultaneously in three beams and all of the other FRBs were detected
in only a single beam. A strong signal appearing in all 13 beams
simultaneously showing a dispersive delay is considered to have a
terrestrial or instrumental origin (called perytons,
\citealt{burkespolaorpery2011}). The origin of these events has been
recently identified as coming from a microwave oven when its door is
opened prematurely and if the telescope is at an appropriate relative angle at
that time \citep{petroffpery}. We did not detect any such events
in our analysis. We found a number of bursts with high S/N in all 13
beams but no dispersive delay in the frequency versus time plot. These events are 
sources of RFI which have near earth origin and are only active for a brief period of time. Nearly $10\%$
of the data show S/N greater than six in more than five beams but do not show dispersive delay. We
did not consider these candidates further in our analysis.
\subsubsection{Manual inspection}

\begin{figure*}
\centering
\includegraphics[scale=0.37]{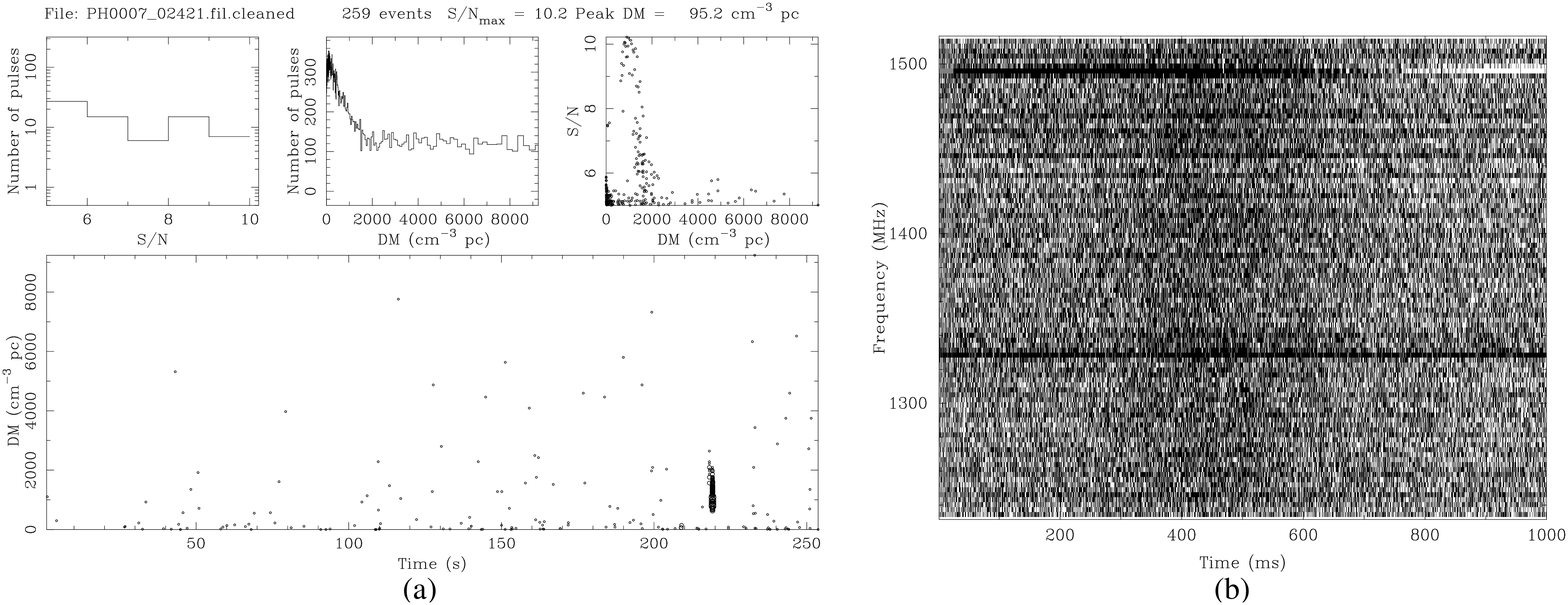}
         
   \caption{The left panel shows a non-astrophysical dispersed burst as seen in
     the lower plot and a corresponding peak in the S/N versus DM plot. 
     The non-dedispersed data is then plotted corresponding to time of the peak
       for one second as seen in the right panel here,
     confirming that it is a false detection.} 
     \label{fig:falsefrb}
    
\end{figure*}

The diagnostic plots obtained after applying all the detection
criteria were manually examined to look for a strong signal corresponding to a peak in the S/N versus DM plot. Such candidates are shortlisted and the detection is confirmed if the signal shows a sweep from
high to low frequency across the observing bandwidth in the frequency versus time plot. Brightness was not a criterion for being shortlisted and some candidates were confirmed as pulsars despite not being detectable in this plot. For FRBs, if a dispersive sweep is absent then it is deemed to be a ``false detection'' (see, e.g., Fig.~\ref{fig:falsefrb}). A burst-like
event was seen in one of the beams (see a bright pulse in the lower plot of 
Fig.~\ref{fig:falsefrb} and a corresponding peak in the S/N versus DM plot in the upper right plot) but no dispersed signal was detected in the frequency versus time plot (see right panel of Fig.~\ref{fig:falsefrb}).
\section{Results}

As summarized in Table~\ref{table:surveyfile}, our single-pulse
search resulted in the detection of 20 of the 42 pulsars detected 
in the original periodicity search by \citet{burgay2006}.
In addition, one RRAT not reported by \citet{burgay2006} was detected.
The discovery of this source, PSR J0410$-$3113, was reported by 
\citet{burkeIII2011} during a single-pulse search on the data
obtained from the high-latitude HTRU survey. In our analysis, only one pulse was detected in this observation for PSR J0410$-$3113, consistent with the non-detection in the periodicity search \footnote{The ATNF pulsar catalog can be accessed online at http://www.atnf.csiro.au/research/pulsar/psrcat}.

\begin{table*}
	\begin{center}
	\caption{All pulsars detected and discovered in the PH survey. Columns 1 to 5 report the pulsar name, Galactic
          longitude and latitude, spin period, and DM, all obtained from the ATNF pulsar
          catalog \citep{atnf2005}. For those that were 
          detected in our single-pulse search method, columns 6, 7, 8, and 9 report the DM obtained in
          this analysis, peak S/N from single-pulse search, number of
          pulses, and width. Column 10 lists S/N from the periodicity
          search obtained from \citet{burgay2006} and column 11 lists the intermittency measure.The RRAT discovered
          by \citet{burkeIII2011} and re-detected in our analysis is denoted by *.}
        \begin{tabular}{lrrrrrrrrrr}
		\hline
		Name & \multicolumn{1}{c}{$l$} & \multicolumn{1}{c}{$b$} & \multicolumn{1}{c}{$P$} & \multicolumn{1}{c}{$DM$} & \multicolumn{1}{c}{$DM_{\rm obs}$} & \multicolumn{1}{c}{($S/N)_{\rm SP}$} & \multicolumn{1}{c}{$N_{\rm pulses}$} & \multicolumn{1}{c}{$W$} & \multicolumn{1}{c}{($S/N)_{\rm per}$} & $r$ \\
	PSR & \multicolumn{1}{c}{($\degree$)} & \multicolumn{1}{c}{($\degree$)} & \multicolumn{1}{c}{($\mathrm{ms}$)} & \multicolumn{1}{c}{($\mathrm{pc~cm^{-3}}$)} & \multicolumn{1}{c}{($\mathrm{pc~cm^{-3}}$)} &  &  & \multicolumn{1}{c}{($\mathrm{ms}$)} & & \\ 
		\hline
		J0343$-$3000 & 227.76 & $-$52.34 & 2597.02 & 20.2 & 22.3 & 21.5 & 22 & 2.0 & 42.7 & 0.50 \\
		J0410$-$3113* & 253.47 & $-$41.95 & 1837.00 & 9.9 & 9.9 & 9.2 & 1 & 4.0 & -- & -- \\
		J0437$-$4715 & 253.47 & $-$41.95 & 5.70 & 2.6 & 2.6 & 12.8 & 4303 & 0.3 & 603.3 & 0.02 \\
		J0448$-$2749 & 228.43 & $-$37.91 & 450.40 & 26.2 & 26.4 & 10.2 & 6 & 4.0 & 28.5 & 0.38 \\
		J0520$-$2553 & 228.51 & $-$30.53 & 241.60 & 33.7 & -- & -- & -- & -- & 34.0 & -- \\
		\\
		J0610$-$2100 & 227.75 & $-$18.18 & 3.86 & 60.6 & -- & -- & -- & -- & 10.1 & -- \\
		J0630$-$2834 & 237.03 & $-$16.75 & 1244.40 & 34.5 & 35.3 & 20.2 & 63 & 16.0 & 277.4 & 0.07 \\
		J0633$-$2015 & 229.33 & $-$12.95 & 3253.21 & 90.7 & 89.7 & 11.0 & 3 & 8.0 & 16.4 & 0.67 \\
		J0636$-$4549 & 254.55 & $-$21.55 & 1984.59 & 26.3 & -- & -- & -- & -- & 11.2 & -- \\
		J0656$-$2228 & 233.66 & $-$8.98 & 1224.75 & 32.4 & 31.1 & 6.9 & 29 & 0.3 & 21.0 & 0.33 \\
		\\
		J0719$-$2545 & 238.93 & $-$5.83 & 974.72 & 253.9 & -- & -- & -- & -- & 20.7 & -- \\
		J0726$-$2612 & 240.08 & $-$4.64 & 3442.31 & 69.4 & 68.8 & 16.8 & 9 & 4.0 & 15.1 & 1.11 \\
		J0729$-$1448 & 230.46 & 1.44 & 251.60 & 92.3 & -- & -- & -- & -- & 32.5 & -- \\
		J0729$-$1836 & 233.83 & $-$0.33 & 510.10 & 61.2 & 61.0 & 7.4 & 26 & 0.3 & 59.3 & 0.13 \\
		J0737$-$3039A & 245.24 & $-$4.50 & 22.70 & 48.9 & -- & -- & -- & -- & 18.7 & -- \\
		\\
		J0737$-$3039B & 245.24 & $-$4.50 & 2773.46 & 48.9 & -- & -- & -- & -- & -- & -- \\
		J0738$-$4042 & 254.27 & $-$9.1 & 374.90 & 160.8 & 161.3 & 20.8 & 644 & 2.0 & 542.1 & 0.04 \\
		J0742$-$2822 & 243.85 & $-$2.43 & 166.70 & 73.8 & 73.0 & 18.3 & 2130 & 2.0 & 63.1 & 0.29 \\
		J0746$-$4529 & 259.20 & $-$10.10 & 2791.03 & 134.6 & -- & -- & -- & -- & 10.3 & -- \\
		J0749$-$4247 & 257.14 & $-$8.33 & 1095.40 & 104.5 & -- & -- & -- & -- & 17.6 & -- \\
		\\
		J0758$-$1528 & 234.54 & 7.24 & 682.20 & 63.3 & 62.9 & 13.4 & 67 & 8.0 & 99.0 & 0.14 \\
		J0818$-$3232 & 251.36 & 1.87 & 2161.26 & 131.8 & -- & -- & -- & -- & 27.3 & -- \\
		J0820$-$1350 & 235.96 & 12.61 & 1238.10 & 40.9 & 39.9 & 19.0 & 28 & 16.0 & 112.5 & 0.17 \\
		J0820$-$3921 & 257.26 & $-$1.58 & 1073.57 & 179.4 & -- & -- & -- & -- & 13.4 & -- \\
		J0820$-$4114 & 258.82 & $-$2.72 & 545.40 & 113.4 & -- & -- & -- & -- & 42.6 & -- \\
		\\
		J0821$-$4221 & 259.83 & $-$3.14 & 396.73 & 270.6 & -- & -- & -- & -- & 10.8 & -- \\
		J0823$+$0159 & 222.06 & 21.26 & 864.80 & 23.7 & 22.3 & 15.8 & 54 & 8.0 & 13.1 & 1.21 \\
		J0828$-$3417 & 254.04 & 2.58 & 1848.90 & 52.2 & 52.5 & 13.3 & 5 &  & 49.2 & 0.27 \\
		J0835$-$3707 & 257.15 & 2.00 & 541.40 & 112.3 & 113.5 & 6.9 & 1 & 8.0 & 33.2 & 0.21 \\
		J0837$-$4135 & 260.98 & $-$0.32 & 751.60 & 147.3 & 147.8 & 23.2 & 303 & 4.0 & 152.2 & 0.15 \\
		\\
		J0838$-$2621 & 248.81 & 8.98 & 308.58 & 116.9 & -- & -- & -- & -- & 9.6 & -- \\
		J0843$+$0719 & 219.40 & 28.22 & 1365.86 & 36.6 & -- & -- & -- & -- & 15.9 & -- \\
		J0846$-$3533 & 257.26 & 4.72 & 1116.00 & 94.1 & -- & -- & -- & -- & 103.1 & -- \\
		J0855$-$3331 & 256.92 & 7.53 & 1267.50 & 86.6 & 89.7 & 9.9 & 12 & 8.0 & 42.5 & 0.23 \\
                J0900$-$3144 & 256.16 & 9.49 & 11.11 & 75.7 & -- & -- & -- & -- & 20.8 & -- \\
                \\
                J0908$-$1739 & 246.19 & 19.86 & 401.60 & 15.8 & -- & -- & -- & -- & 23.9 & -- \\
                J0922$+$0638 & 225.48 & 36.40 & 430.60 & 27.3 & 26.4 & 18.2 & 195 & 4.0 & 96.8 & 0.19 \\
                J0944$-$1354 & 249.20 & 28.86 & 570.20 & 12.4 & -- & -- & -- & -- & 43.1 & -- \\
                J0953$+$0755 & 228.97 & 43.71 & 253.00 & 2.9 & 3.0 & 14.0 & 657 & 2.0 & 128.7 & 0.11 \\
                J1022$+$1001 & 231.86 & 51.11 & 16.40 & 10.3 & 10.4 & 9.6 & 266 & 0.5 & 318.7 & 0.03 \\
                \\
                J1024$-$0719 & 251.77 & 40.53 & 5.10 & 6.4 & -- & -- & -- & -- & 33.4 & -- \\
                J1038$+$0032 & 247.15 & 48.47 & 28.85 & 26.6 & -- & -- & -- & -- & 18.1 & -- \\
                J1046$+$0304 & 246.48 & 51.71 & 326.20 & 25.3 & 28.2 & 9.8 & 1 & 8.0 & 25.5 & 0.38 \\     
		\hline
		\end{tabular}

		\label{table:surveyfile}
		
	\end{center}
\end{table*}

\subsection{Energy measurements}

\begin{figure*}
    \includegraphics[]{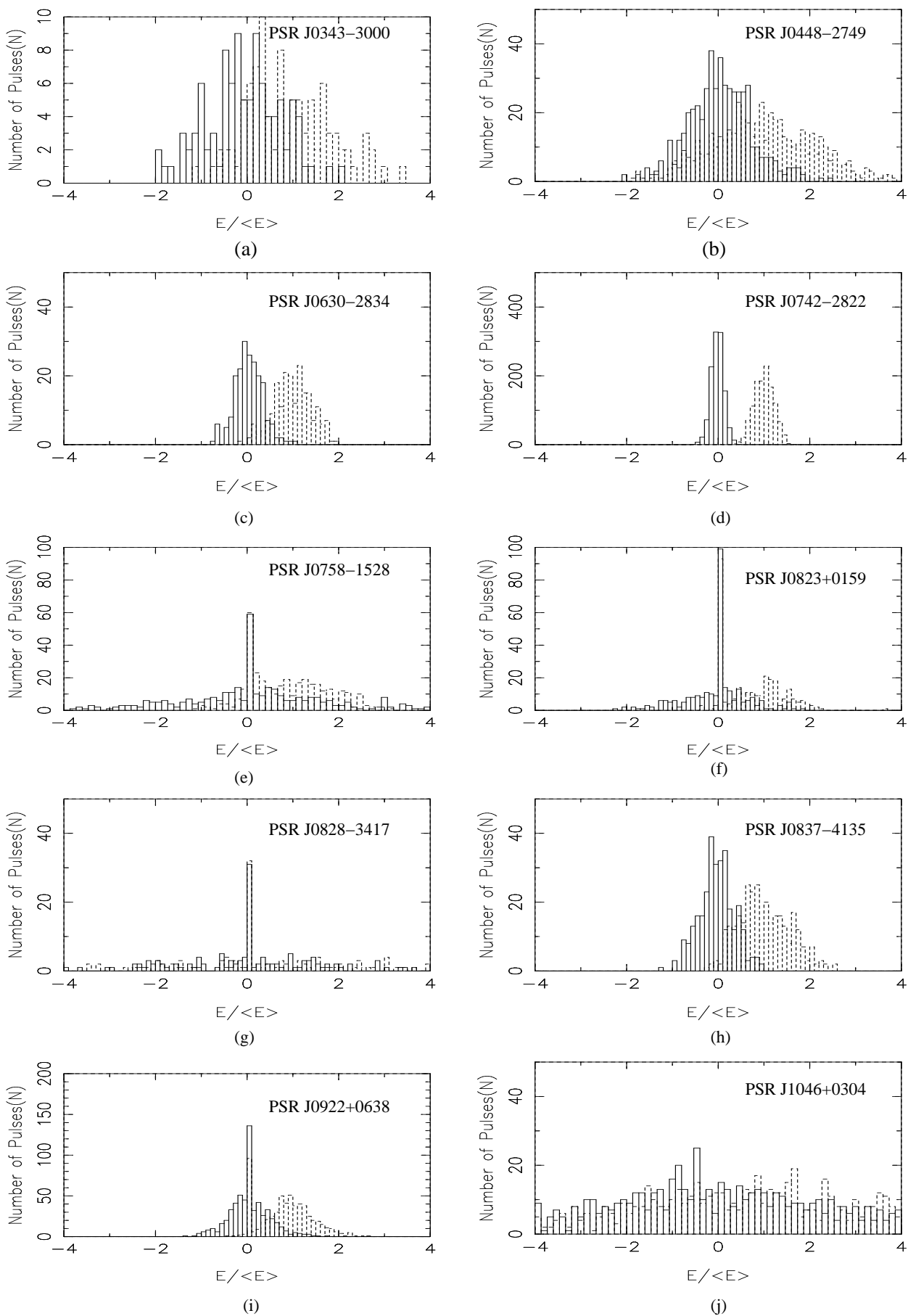}
  \caption{Normalized histograms of on-pulse (dashed) and 
off-pulse (solid) energies, for 10 of the total 
21 re-detected pulsars.}\label{fig:energyhistograms}
\end{figure*}

Integrated profiles for each pulsar were obtained by folding the
dedispersed time series using ephemerides available from the ATNF
pulsar catalog \citep{atnf2005}. To construct the pulse energy histograms, the procedure  described in \citet{ritchings1976} is followed. The position and widths of the on-pulse and off-pulse
windows were determined by visual inspection of the integrated pulsar
profile. Baseline estimation was done using the off-pulse bins, and was
subtracted from the data for each pulse. The total energy for each pulse in the on-pulse window was calculated and is scaled to account for different widths of on and off windows.. The on-pulse data were taken in blocks of about 100 pulses. The on-pulse and off-pulse
normalized energies, $E/\bar{E}$, were calculated for each block by
dividing on-pulse and off-pulse energies within the block by the mean on-pulse
energy of that block to account for variations due to interstellar
scintillation. The pulse energy histograms were constructed for 10 of
the total 17 re-detected pulsars, as shown in
Fig.~\ref{fig:energyhistograms}, based on the number of pulses detected 
during the total observation time. We did not create energy distribution plots for sources
with fewer than 100 pulses per observation.

For some pulsars, the on-pulse distribution peaks at the mean energy, as seen for
PSR~J0742$-$2822, PSR~J0630$-$2834, PSR~J0837$-$4135, and
PSR~J0922$+$0638. In the case of PSR~J0742$-$2822, the energy
histograms separate out clearly with no zero energy excess in the ON
histogram, suggesting that this pulsar does not null. For other pulsars (PSR~J0448$-$2749,
PSR~J0828$-$3417, and PSR~J1046+0304),
the histograms overlap just because the S/N is low and not necessarily because the pulsar
actually nulls.  But it is important to note here
that the statistics are limited by the relatively short observation
time in the survey. So, we did not fit any Gaussians to the histograms since the errors would be large.
Even if the pulsar is nulling, there are
insufficient pulses to form a distribution for estimation of the
nulling fraction in our analysis.

\subsection{Intermittency measure}

\begin{figure} 	
 \includegraphics[width=6.0cm,angle=270]{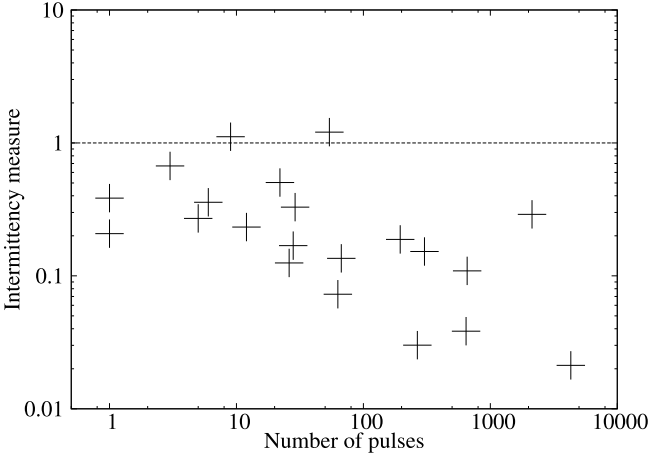}
  \caption{Intermittency measure for each pulsar detection in our survey.} 
\label{fig:interph}
 \end{figure}

The two search algorithms used for pulsar searching show varying
levels of efficiency which depend upon the properties of each particular
pulsar \citep{mccordes2003}. Following \citet{deneva2009}, we
calculated the intermittency ratio 
\begin{equation}
r = \frac{(S/N)_{\rm SP}}{(S/N)_{\rm per}}
\end{equation}
for each pulsar from the S/N value of the single-pulse and
periodicity searches and \textit{r} is plotted versus the number of periods in $\mathrm{T_{obs}}$ (see 
Fig.~\ref{fig:interph}). All the pulsars re-detected in our analysis
except PSR J0410$-$3113 were detected in a previous periodicity
search. The pulsars on the upper left have longer periods and the
pulsars on the lower right of this plot are millisecond
pulsars. Pulsars with $\textit{r} > 1$ (PSRs J0726$-$2612 and J0823$+$0159) are more
likely to be detected with single-pulse searches. PSR J0726$-$2612,
with $\textit{r} \sim 1.1$, has $P = 3.4~\mathrm{s}$ and DM~$= 69~\mathrm{pc~cm^{-3}}$ 
and PSR~J0437$-$4715, with $\textit{r} \sim 0.02$, has $P =
5.7~\mathrm{ms}$ and DM~$ = 2.6~ \mathrm{pc~cm^{-3}}$.
These ratios are higher than one would expect in longer surveys, because of the dependence
of sensitivity on number of pulses. The integration time for this survey was relatively 
short (265~s) and we detected $50\%$ of pulsars from single-pulse search. This can be compared to the PM
survey with long integration time (2100~s), in which the
single-pulse detections were only about $30\%$ of the total pulsar
discoveries from periodicity search \citep{keane2010}.

\section{Event rate of FRBs}\label{sec:rate}
\begin{figure}
\includegraphics[width=6.0cm, angle=270]{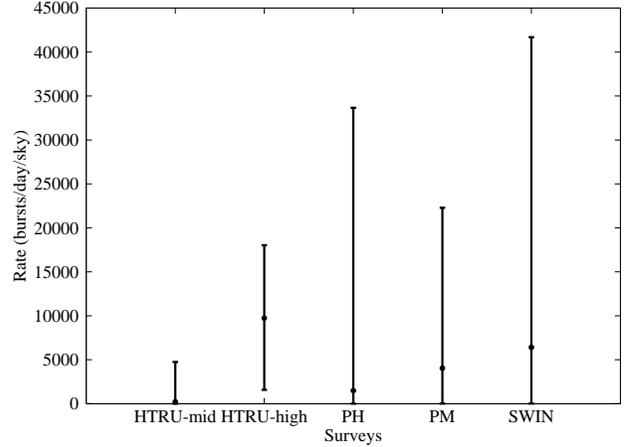}
\caption{The modified FRB event rates based on individual survey results and corrected for detection sensitivity based on the search algorithms and the backend used.}
\label{fig:ratesall}  
\end{figure}

\begin{figure}
\includegraphics[width=9.0cm]{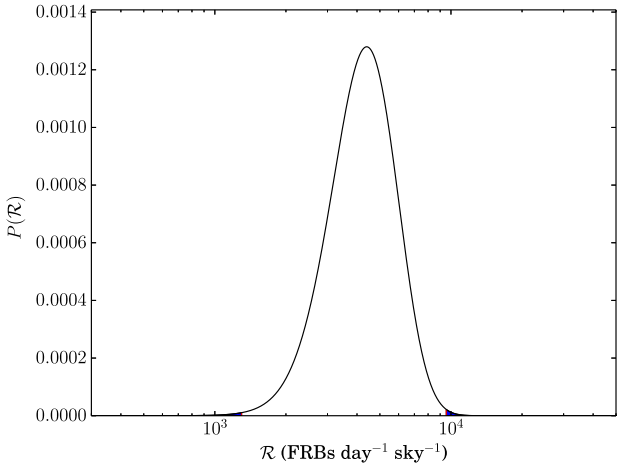}
\caption{The combined posterior PDF of the FRB event rate taking into
  account all the published Parkes survey results carried out to date
  and correcting for detection sensitivity based on the search
  algorithms used and the backend used. The all-sky event rate for sources with a fluence above 4.0~Jy ms is ${\cal R} =
4.4^{+5.2}_{-3.1} \times 10^3$~FRBs~day$^{-1}$~sky$^{-1}$ .}
\label{fig:posteriorall}
\end{figure}

Although our analysis did not result in any new FRB
detections, it is important to consider the impact of this null
result on the FRB event rate, ${\cal R}$. To put our results into context,
we also consider a number of other surveys at Parkes 
in the calculation below. To constrain ${\cal R}$, we apply a Bayesian
approach which uses the FRB detections in each survey. Using
Bayes' theorem (see, e.g., \citealt{walljenkins}),
the posterior probability density function for ${\cal R}$, given
the detection of $n$ pulses
\begin{equation}
\label{eq:bayes}
p({\cal R}|n) \propto p(n|{\cal R})p(\cal R),
\end{equation}
where $p(n|\cal R)$ is the likelihood of getting $n$ detections given
some $\cal R$ and $p(\cal R)$ is the prior on the rate of FRBs taken
to be uniform. Taking $n = 0$ from \citet{petroffhtru2014} and assuming the counting of 
these rare events as a Poisson process, the likelihood function
\begin{equation}
p(0|{\cal R}) = \frac{({\cal R} A_{P}T_{P})^{0} \exp(-fRT_{P}A_{P})}{0!},
\end{equation}
where $A_{\rm P}=4449$~deg$^2$ and $T_{\rm P}=540$~sec is {\bf the} total
observation time for each pointing. The subscript $P$ corresponds to
the values for Parkes. The
numerical factor $f = (86400 \times 41253)^{-1}$ is inserted to
compute $\cal R$ in units of bursts~day$^{-1}$~sky$^{-1}$. The
posterior probability density of the FRB event rate can be computed as
\begin{equation}
\label{eq:posterior}
p({\cal R}|0) = {K_1} {\cal R}^0 \exp(-fRT_{P}A_{P}), 
\end{equation}
where ${K_1}$ is a normalizing constant which ensures that the above
expression integrates over all values of ${\cal R}$ to unity.
Integrating this function, we find that the mean FRB rate based on
zero FRBs in the HTRU mid-latitude survey is ${\cal
  R}=0.22^{+4.5}_{-0.21}\times10^3$~FRBs~day$^{-1}$~sky$^{-1}$, where
the uncertainties represent the $99\%$ confidence interval. To include the FRB detections and null results from subsequent
surveys, a similar calculation is 
performed to determine the FRB event rates for individual surveys as listed in Table 2. 
The surveys were processed using different search algorithms and therefore the rates need to be modified 
following the results of \cite{keanepetroff2015} in order to combine the individual event rates. As per the response curve shown in Figure 1 of \cite{keanepetroff2015}, we get 
the corrected S/N from the use of these algorithms by defining the efficiency factors,
$\eta_{\rm{ddall}}$ and $\eta_{\rm{seek}}$. For {\tt dedisperse\textunderscore all}, we find
\begin{equation}
\eta_{\rm{ddall}}=\frac{S/N_{\rm \tt dedisperse\textunderscore all}}{S/N_{\rm max}}  = 0.83.
\end{equation} 
and, for {\tt seek},
\begin{equation}
\eta_{\rm{seek}}=\frac{S/N_{\rm seek}}{S/N_{\rm max}} = 0.89.
\end{equation}
 The FRB searches we consider in this analysis are the results of reprocessing the Parkes Multibeam (PMPS)
survey (\citet{keane2010,keane2011}, used {\tt destroy}), the PH survey (this paper, used {\tt seek}), the
Swinburne intermediate and high-latitude (SWIN) surveys
(\citet{edwards2001}, \citet{jacoby2009}, \citet{burkespolaorfrb2014}, used {\tt dedisperse\textunderscore all}), the HTRU high-latitude (HTRU high) survey (\citet{thorntonfrb2013}, \citet{thornton2013}, used {\tt dedisperse\textunderscore all}), and the HTRU mid-latitude (HTRU mid) survey
(\citet{petrofffrb}, used {\tt Heimdall}). In addition to this, the HTRU surveys were carried out using the digital back-end, Berkeley-Princeton-Swinburne Recorder (BPSR), whereas the older Parkes surveys used the analogue filterbank (AFB). In order to compare all Parkes surveys, the digitization loss factors ($\beta$) depending on the back-end used need to be considered in our analysis. For AFB, $\beta=1.25$ and for BPSR, $\beta=1.07$ \citep{kou2001}. We then insert the efficiency factors into the minimum fluence equation calculated for each processed survey
\begin{equation} 
F_{\rm min} = \frac{\beta~T_{\rm sys}(S/N)_{\rm min}}{G~\eta_{\rm soft}\sqrt{N_{\rm p} \Delta f}}\sqrt{w},
\end{equation} 
where $T_{\rm sys}$ is the system temperature, $G=0.66~{\rm K/Jy}$ is the telescope gain, $N_{\rm p}$ is number of polarizations, $\Delta f$ is the bandwidth, $\eta_{\rm soft}$ is either $\eta_{\rm seek}$ or $\eta_{\rm ddall}$ depending the survey, and $w$ is the pulse width equal to the sampling time when calculating the minimum fluence. The HTRU mid-latitude survey has the lowest minimum fluence as can be seen in Table~\ref{table:rates}. The event rates of other surveys are scaled to this lowest $F_{\rm min}^{-3/2}$ which is the simplest model assuming a uniform distribution of standard-candle FRBs in Euclidean geometry. The modified event rates are shown in Fig.~\ref{fig:ratesall}. Combining all these individual rate
estimates, Fig.~\ref{fig:posteriorall} shows our current best estimate
of ${\cal R}$. The mean FRB rate from this distribution is
$4.4^{+5.2}_{-3.1} \times 10^3$~FRBs~day$^{-1}$~sky$^{-1}$ for sources with
a fluence above 4.0~Jy ms at 1.4~GHz, where the
uncertainties represent a $99\%$ confidence interval. To demonstrate
that this rate is consistent with all the surveys considered here, we
list in Table~\ref{table:rates} the predicted upper and lower bounds
on the number of FRBs expected in each survey which use these $99\%$
confidence intervals on ${\cal R}$ and scale it back to each survey's fluence
limit. In addition, we make predictions for future
analyses of the Perseus Arm (PA) pulsar survey \citep{burgay2013} and
HTRU low-latitude survey \citep{thornton2013}.

The event rate is estimated assuming FRBs are 
uniformly distributed on the sky.
 We now consider the impact of this
assumption. The Galactic effects such as dispersion in the ISM, scattering in the
ISM, scintillation, and sky temperature can limit the sensitivity of
a survey. Following the discussion in \citet{petroffhtru2014} about
decreased sensitivity to FRBs at $|b| \leq 15 \degree$, we compare
the sensitivity based on sky coverage for the Parkes surveys
considered in our calculations. The non-Galactic DM contribution at
high latitudes range between 520 and 1070~$\mathrm{~cm^{-3}~pc}$. 
The Galactic dispersion at these latitudes is only about
50~cm$^{-3}$~pc, whereas the average Galactic dispersion at intermediate
latitudes and near the Galactic center are
380~cm$^{-3}$~pc and 1780~cm$^{-3}$~pc respectively. The FRB pulses
with an additional non-Galactic DM contribution at these lower latitudes would still be
recovered in the surveys considered above as they have been searched
to a sufficient DM, with the maximum trial DM in each of these surveys
being $> 2000$~cm$^{-3}$~pc. 

The average sky temperature values for four of these surveys
range between 0.85~K to 3.18~K and about 6.14~K for the PMPS survey
\citep{burkespolaorfrb2014}. Sky temperature is therefore not a
significant factor in limiting the sensitivity when comparing these surveys.

FRBs discovered so far (except
FRB~110220) show few effects of scattering \citep[see,
  e.g.,][]{lkmj2013}. \citet{petroffhtru2014} determined that more than $85\%$ of
survey pointings in the intermediate latitude survey are still
sensitive to FRB signals even in the presence of strong scattering in
the ISM near the Galactic center. This percentage will differ for AFB surveys as the number of channels, 
sampling time, and digitization factors are different but is still small compared to overall uncertainty in the FRB event rate at this point.

\citet{petroffhtru2014} analyzed how the combination of these effects
limits sensitivity for survey pointings and determined that a
simulated FRB pulse with properties similar to the FRBs in
\citet{thornton2013} falls below the detection threshold in only
$14\%$ of all intermediate latitude pointings assuming no scattering
in the ISM. We combine the individual
rates since the sensitivity variations between AFB and BPSR are within the
uncertainty of rate of FRBs. \citet{petroffhtru2014} argue that since
the percentage of pointings no longer sensitive to FRB pulses
decreases to $14\%$, the null result is discrepant with the original
predictions based on a higher event rate. Although, if the event rate
of FRBs is much lower than previous estimates, it still explains the
null result at intermediate latitudes (Table 2). Therefore, we argue that the
lack of detections at intermediate latitudes and the null result in our analysis of a high-latitude
survey are likely to be due to a lower FRB rate and does not necessarily imply a dearth of FRBs at intermediate
latitudes.

\begin{table*}
	\begin{center}
	\caption{Survey parameters, FRB detections, event rates and predicted
         numbers of FRBs for multibeam pulsar surveys carried out at Parkes.
         The survey abbreviations are listed in the text. For each survey, we
         provide the dwell time ($T$), area covered ($A$), bandwidth ($\Delta f$), digitization factor ($\beta$), efficiency factor ($\eta$), sampling time ($t_{\rm samp}$), system temperature ($T_{\rm sys}$), minimum fluence ($F_{\rm min}$), number of FRBs
         found ($N_{\rm FRBs}$), estimated FRB event rate (${\cal R}$), and the modified FRB event rate ($\cal R_{\rm mod}$). 
         Using the combined FRB event rate, we give the
         number of FRBs expected to be found ($N_{\rm expected}$), as well
         as lower and upper bounds on this expected number ($N_{\rm lower}$ and
         $N_{\rm upper}$).
         }
          \begin{tabular}{lrrrrrrr}
		\hline
		\multicolumn{1}{c}{Parameter} & \multicolumn{1}{c}{HTRU High} & \multicolumn{1}{c}{HTRU Mid} & \multicolumn{1}{c}{HTRU Low} & \multicolumn{1}{c}{SWIN} & \multicolumn{1}{c}{PA} & \multicolumn{1}{c}{PMPS} & \multicolumn{1}{c}{PH} \\
		\hline		
Dwell time $T$ (s) & 270 & 540 & 4300 & 265 & 2100 & 2100 & 265 \\
Area $A$ (deg$^2$) & 21671.6 & 4448.8 & 769.5 & 7264.3 & 599.3 & 2097.4 & 3979.5 \\
Bandwidth $\Delta f$ (MHz) & 400 & 400 & 400 & 288 & 288 & 288 & 288 \\
Digitization $\beta$ & 1.07 & 1.07 & 1.07 & 1.25 & 1.25 & 1.25 & 1.25 \\
Efficiency $\eta_{\rm soft}$ & 0.83 & 1.0 & 1.0 & 0.83 & 1.0 & 1.0 & 0.75 \\
Sampling time $t_{\rm samp}$ (ms) & 0.064 & 0.064 & 0.064 & 0.125 & 0.125 & 0.25 & 0.125 \\
System temperature $T_{\rm sys}$ (K) & 29.0 & 30.5 & 35.5 & 29.5 & 29.2 & 33.6 & 29.0 \\
Fluence $F_{\rm min}$ (Jy ms) & 4.6 & 4.0 & 4.6 & 8.9 & 7.3 & 11.9 & 8.2 \\
   $N_{\rm FRBs}$ & 5 & 0 & - & 1 & - & 1 & 0 \\
${\cal R}$ ($10^3$~FRBs~d$^{-1}$~sky$^{-1}$) & $7.9^{+14.6}_{-6.6}$ & $0.22^{+4.5}_{-0.21}$ & $ - $ & $1.9^{+10.2}_{-1.8}$ & $ - $ & $0.81^{+4.5}_{-0.8}$ & $0.5^{+10.1}_{-0.4}$ \\
${\cal R}_{\rm mod}$ ($10^3$~FRBs~d$^{-1}$~sky$^{-1}$) & $9.7^{+18.0}_{-8.2}$ & $0.22^{+4.5}_{-0.21}$ & $ - $ & $6.2^{+34.1}_{-6.2}$ & $ - $ & $4.1^{+22.3}_{-4.0}$ & $1.5^{+33.7}_{-1.4}$ \\
\midrule
   $N_{\rm expected}$ & 2 & 3 & 3 & 1 & 1 & 1 & 0 \\  
   $N_{\rm upper}$ & 5 & 6 & 7 & 2 & 1 & 2 & 1 \\
   $N_{\rm lower}$ & 1 & 1 & 1 & 0 & 0 & 0 & 0 \\
  	\bottomrule
		\end{tabular}
		\label{table:rates}
		
	\end{center}
\end{table*}

\section{Conclusions}

We have presented the results of a single-pulse search of the PH
pulsar survey. We re-detected 20 of the previously known pulsars
reported by \citet{burgay2006} and detected one RRAT. Out of these 21
re-detections, we have constructed pulse energy histograms for 10
pulsars for which the observed number of pulses was more than 100
. For PSR J0742$-$2822, the on-pulse and off-pulse
histograms separate out clearly, suggesting it does not null over the observing
span. The observation time, being relatively short, is insufficient for
estimation of the nulling fraction. We demonstrated, using a Bayesian
approach that the lack of FRB detections, and detection rates in other
surveys, is consistent with an all-sky FRB event rate ${\cal R} =
4.4^{+5.2}_{-3.1} \times 10^3$~FRBs~day$^{-1}$~sky$^{-1}$, for sources with
a fluence above 4.0~Jy ms at 1.4~GHz, where the
uncertainties represent a $99\%$ confidence interval. This event rate
takes into account the decrease in detection sensitivity as a result
of the search algorithms used in the analysis and the different backends used in these surveys.
We argue that previous
suggestions of a dearth of FRBs at intermediate Galactic latitudes is
likely to be a result of the assumption of a higher event rate. The
revised event rate found here is in good agreement with 
\citet{keanepetroff2015} who estimated 
${\cal R} \sim 2500$~FRBs~day$^{-1}$~sky$^{-1}$. We concur with these
authors that a larger sample of FRBs is required to more meaningfully
constrain the rate.

\section{acknowledgements}
This work was partially supported by the Astronomy and Astrophysics Division of the National Science Foundation via grant AST-1309815. 
We thank the referee for helpful comments on the manuscript.

\section{Appendix}
\citet{levin2012} describes the the algorithm which accounts for the amount of pulse broadening
caused by the size of the previous DM step and then determines the next trial DM.
The appropriate step size is computed from the effective pulse width ($w_{\rm eff}$), which is a quadrature sum of the intrinsic pulse width
($t_{\rm in}$), the smearing due to scattering ($t_{\rm scatt}$),
dispersion smearing within each frequency channel ($t_{\rm DM}$), sampling time of a particular
survey ($t_{\rm samp}$), and smearing across all frequency channels due to
the dispersion measure step size ($t_{\rm \Delta DM}$)
\begin{equation}
w_{\rm eff} = \sqrt{{t_{\rm in}^2 + t_{\rm scatt}^2 + t_{\rm samp}^2 + t_{\rm DM}^2 + t_{\Delta{\rm DM}}^2}},
\end{equation} 
where, for frequencies in GHz and DM in units of cm$^{-3}$~pc,
\begin{equation}
t_{\rm DM} = 8300 {\rm s} \, \left(\frac{\Delta f~\rm DM}{f^3}\right)
\end{equation}
and 
\begin{equation}
t_{\rm \Delta DM} = 8300 {\rm s} \, \left(\frac{n_{\rm chan} \Delta f~{\rm \Delta DM}}{4f^3}\right).
\end{equation}
Here $n_{\rm chan}$ is the number of frequency channels and $\Delta \rm{DM} =	
\rm{DM} - \rm{DM}^\prime$. Since 4 samples are packed per 64-bit word, the equation is divided by a factor of 4. Since the total pulse width smearing of $25\%$ is chosen, the pulse broadening fraction due to the DM step is $\epsilon = 1.25$. Then the effective pulse width at the new trial DM, with respect to the last trial DM$^\prime$ is 
\begin{equation}
\left(w_{\rm eff| DM}\right)= \epsilon~\left(w_{\rm eff| DM^\prime}\right)
\end{equation}
Solving for DM by equating $w_{\rm eff}$ at the new trial value, DM, with respect to the last trial value, DM$^\prime$, 
\begin{equation}
\rm{DM} = \frac{b^2~\rm{DM}^\prime + \sqrt{-a^2~b^2\rm{DM}^{\prime 2} + a^2c + b^2c}}{a^2 + b^2}, 
\end{equation}
where 
\begin{equation}
a = 8300~{\rm s}~ \frac{\Delta f~{\rm DM}}{f^3},
\end{equation}
\begin{equation}
b = t_{\rm \Delta DM} = 8300~{\rm s}~ \frac{n_{\rm chan} \Delta f~{\rm \Delta DM}}{4~f^3},
\end{equation}
and
\begin{equation}
c = \epsilon^2~(t_{\rm in}^2 + t_{\rm scatt}^2 + t_{\rm samp}^2 + t_{\rm DM}^2) - t_{\rm samp}^2 - w_{\rm in}^2.
\end{equation}
The DM values are obtained using Equation 14 where DM step is finely
spaced for small values of DM. For higher DM values, pulse smearing is
greater than the sampling time and hence the temporal resolution is
decreased giving DM channels that are spaced more coarsely.
\label{lastpage}

\end{document}